\documentclass[final]{elsarticle}

\usepackage{lineno,hyperref}
\modulolinenumbers[5]
\usepackage{multirow}
\usepackage{makecell}
\usepackage{lipsum}
\usepackage{enumitem}
\usepackage{graphicx}
\usepackage{wrapfig}
\usepackage{subcaption}
\usepackage{xcolor, soul}
\usepackage[normalem]{ulem}
\usepackage[framemethod=tikz]{mdframed}
\usepackage{cancel}
\usepackage{fancyhdr}

\newcommand\hldel{%
 \bgroup
 \expandafter\def\csname sout\space\endcsname{\bgroup \ULdepth =-.8ex \ULset}%
 \markoverwith{\textcolor{red!30}{\rule[-.5ex]{.1pt}{2.5ex}}}%
 \ULon}

\newcommand\hladd{%
 \bgroup
 \expandafter\def\csname sout\space\endcsname{\bgroup \ULdepth =-.8ex \ULset}%
 \markoverwith{\textcolor{green!20}{\rule[-.5ex]{.1pt}{2.5ex}}}%
 \ULon}

\newcommand\hlcom{%
 \bgroup
 \expandafter\def\csname sout\space\endcsname{\bgroup \ULdepth =-.8ex \ULset}%
 \markoverwith{\textcolor{blue!20}{\rule[-.5ex]{.1pt}{2.5ex}}}%
 \ULon}

\journal{International Journal of Human-Computer Studies}

\biboptions{authoryear}

\begin{document}

\begin{frontmatter}
\pagestyle{fancy}
\fancyhead[CO]{\textcolor{blue}{This article was accepted in the International Journal of Human-Computer Studies on January 4, 2022. To access and cite the published article, visit \url{https://doi.org/10.1016/j.ijhcs.2022.102772}.
}}
\fancyhead[R]{}

\title{Modeling and mitigating human annotation errors to design efficient stream processing systems with human-in-the-loop machine learning
}





\author[gmuaddress]{Rahul Pandey\corref{mycorrespondingauthor}}\cortext[mycorrespondingauthor]{Corresponding author}
\ead{rpandey4@gmu.edu}
\ead[url]{http://mason.gmu.edu/~rpandey4/}
\author[gmuaddress]{Hemant Purohit}

\author[upfaddress,icreaaddress]{Carlos Castillo}
\author[wsuaddress]{Valerie L. Shalin}

\address[gmuaddress]{George Mason University, 4400 University Dr, Fairfax, Virginia, USA}
\address[upfaddress]{Universitat Pompeu Fabra, Plaça de la Mercè, 10-12, Barcelona, Spain}
\address[icreaaddress]{ICREA, Pg. Llu{\'i}s Companys 23, Barcelona, Spain}
\address[wsuaddress]{Wright State University, 3640 Colonel Glenn Hwy, Dayton, Ohio, USA}

\begin{abstract}
High-quality human annotations are necessary for creating effective machine learning-driven stream processing systems. We study hybrid stream processing systems based on a Human-In-The-Loop Machine Learning (HITL-ML) paradigm, in which one or many human annotators and an automatic classifier (trained at least partially by the human annotators) label an incoming stream of instances. This is typical of many near-real-time social media analytics and web applications, including annotating social media posts during emergencies by digital volunteer groups. From a practical perspective, low-quality human annotations result in wrong labels for retraining automated classifiers and indirectly contribute to the creation of inaccurate classifiers. 

Considering human annotation as a psychological process allows us to address these limitations. We show that human annotation quality is dependent on the ordering of instances shown to annotators and can be improved by local changes in the instance sequence/order provided to the annotators, yielding a more accurate annotation of the stream. We 
adapt a theoretically-motivated human error framework 
of mistakes and slips
for the human annotation task to study the effect of ordering instances (i.e., an  ``annotation schedule''). Further, we propose an error-avoidance approach to the active learning 
paradigm for stream processing applications 
robust to these likely human errors 
(in the form of slips)
when deciding a human annotation schedule. We 
support the human error framework using crowdsourcing experiments and evaluate the proposed algorithm against standard baselines for active learning via extensive experimentation on classification tasks of filtering relevant social media posts during natural disasters. 

According to these experiments, considering the order in which data instances are presented to a human annotator leads to increased accuracy for machine learning and awareness of the potential properties of human memory for the class concept, which may affect annotation for automated classifiers. Our results allow the design of hybrid stream processing systems based on the HITL-ML paradigm, which requires the same amount of human annotations, but that has fewer human annotation errors. Automated systems that help reduce human annotation errors could benefit several web stream processing applications, including social media analytics and news filtering.
\end{abstract}


\end{frontmatter}

\section{Introduction}
\label{sec:1}
\begin{figure}[h!]
	\centering
	\begin{subfigure}[c]{0.8\linewidth}
		\includegraphics[width=\linewidth, trim={0cm 0 0cm 0}]{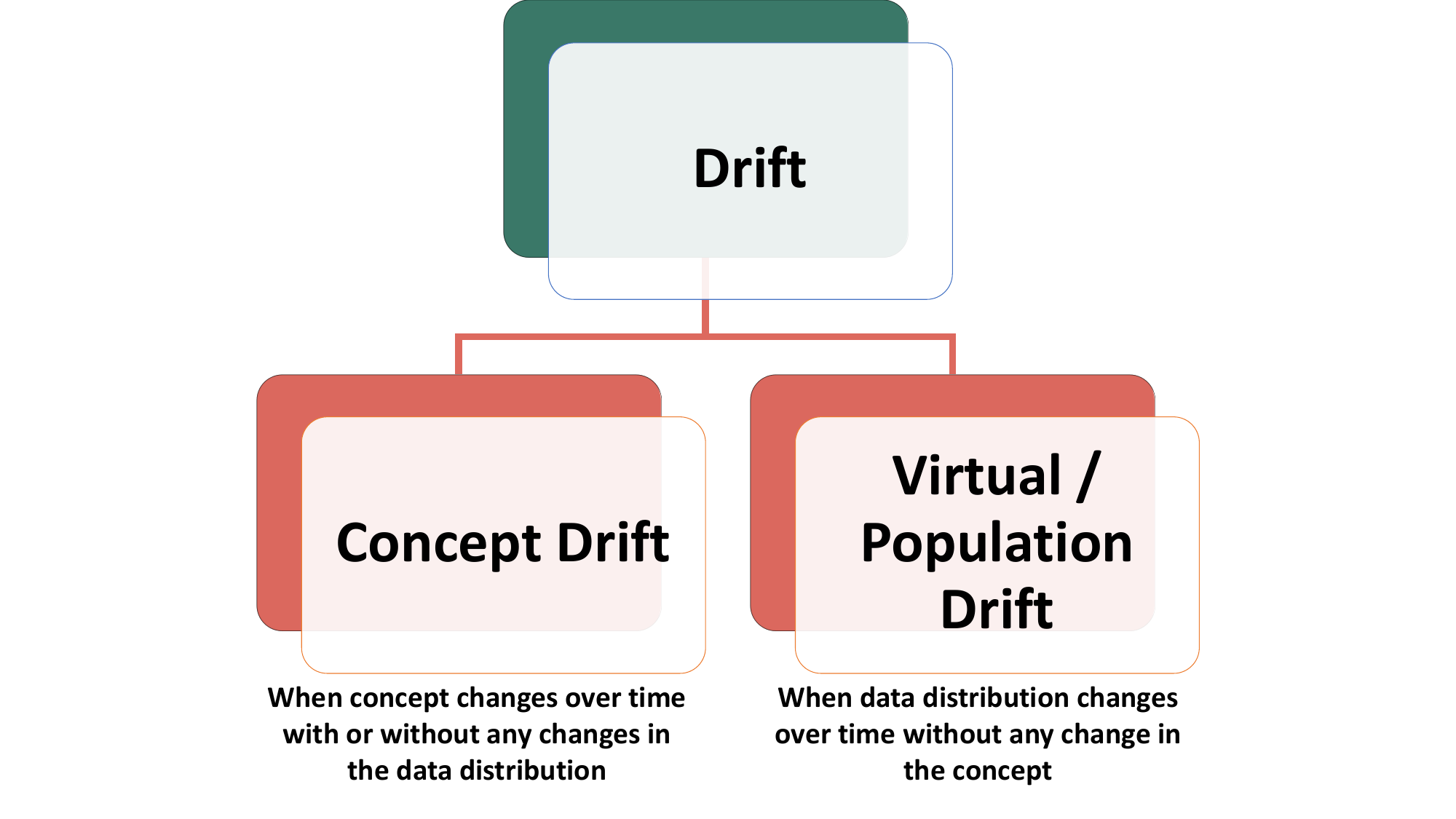}
	\end{subfigure}
  \caption{Categories of drift in streaming data \citep{gama_survey_2014}.}
  \label{fig:drift}
\end{figure}
Filtering high-volume, high-velocity data streams is a typical process in many application domains such as journalism, public health, and crisis management. In this process, an avalanche of data must be filtered  and classified  to prevent recipient information overload and filter failure \citep{shirky2008s}. These continuous streams of data are often noisy, sparse, and redundant. Humans cannot keep pace with the high velocity and volume of data. A purely human-annotation based filtering system does not scale. These data streams are also problematic for purely automated/machine-annotation based filtering systems; depending on the application, they may have limited accuracy. In the case of supervised classifiers for such automated filtering, data sampled from previously collected streams can 
bootstrap classifier training. However, it is invaluable to have annotations on samples specifically from the current data stream to adapt the pre-trained classifier model for the new data. Hence, to achieve high accuracy in this process, online human annotation tasks are needed within an active learning 
paradigm \citep{gama_survey_2014}, sometimes at a large scale. Fortunately, social media and mobile devices have provided an unprecedented opportunity for the public to participate by volunteering in stream processing applications for digital humanitarianism, citizen science purposes, etc. A popular option  for annotating complex data streams has been to create 
hybrid stream processing 
systems 
through a composition of human annotation tasks and automatic online classification \citep{imran2013engineering, lofi_design_2014}. 

In this paper, we study a hybrid online classification setting that categorizes relevant instances from a social media data stream using human annotation tasks and an active learning paradigm. Drawing on both classic 
\citep{ebbinghaus1913memory} 
and contemporary \citep{anderson2000learning} cognitive psychology, we analyze the effective decay related to attentional processes (described below) in human memory in contributing to errors while doing human annotation tasks\footnote{We appreciate the distinction between absent memory traces and the challenges of retrieval \citep{tulving_availability_1966}. For the purposes of this paper, the net result is memory decay that results in effective forgetting.}. 

\paragraph{Data challenges in hybrid stream processing} A key challenge in stream processing is temporal variation in the concept space. This includes changes in the 
distribution of data that leads to change in decision boundaries
(virtual drifts), changes in the 
population from which future samples will be drawn
(population drift), and changes in the definition of a concept (concept drift) \citep{gama_survey_2014} as illustrated in Figure \ref{fig:drift}. For example, consider the task of processing crisis-related instances posted on social media during a natural disaster, such as a hurricane. To find instances that can help emergency managers in a response agency, we need to categorize them as irrelevant or relevant for actionable services \citep{purohit_social-eoc_2018}, and in the case of relevant instances, further categorize them into fine-grained information classes such as infrastructure damage, donations, and so on \citep{castillo_big_2016}. In this setting, both virtual drifts and population drifts occur as the crisis unfolds. An example of virtual drift is content variation as a crisis evolves  \citep{sutton_terse_2015, olteanu_what_2015}. Consider a class concept such as \emph{caution and advice}. In the beginning, instances might be urgent and generic, warning the public about a potentially dangerous event (such as a hurricane warning). Later, the same category of instances may become more specific and less urgent (such as warning people to avoid drinking contaminated water). An example of population drift is change in the prevalence of different class instances, which follow a certain progression across many events \citep{olteanu_what_2015}. For instance,  immediately after a sudden onset crisis event, instances of caution and advice appear. Later on, other classes of information may be prevalent such as appeals for relief donations. These temporal variations are expected. They have a potential effect on 
annotation quality due to the learning behavior of human annotators about the representation of a class concept, which, in turn, impacts the entire system when used to train the automatic part of a hybrid system.

\paragraph{Human challenges in hybrid stream processing} Human factors in the annotation process affect the quality of annotations for hybrid stream processing systems. Systems that rely on some form of crowdsourcing are affected by cognitive properties of human annotators, including their attentional heuristics (e.g., the fit with prior experience, the associated positive or negative affect) and vigilance (the 
ability to sustain high attention over time) \citep{burghardt_quantifying_2018}. High mental workload (e.g., demands on inference and decision making) causes a deterioration in annotation quality, known as \emph{annotator burnout} \citep{marshall_experiences_2013}, which can cause increased fatigue and reduced motivation to maintain accuracy. To prevent annotator burnout, one may cap the maximum number of annotation tasks per unit of time that the annotator must perform, which can reduce workload \citep{purohit_ranking_2018}. Nevertheless, human error persists in the execution of annotation tasks.

Psychologists distinguish between two types of human error: mistakes and slips \citep{reason_human_2000}. Mistakes result from incorrect or incomplete knowledge \citep{reason_human_2000}. In the annotation task context, this corresponds to annotators who have not yet grasped the concept to be annotated or who are annotating new instances for which they have not yet acquired a correct representation. Slips are errors in the presence of correct and complete knowledge \citep{reason_human_2000, norman1981categorization}, i.e., annotator knowledge is correct, but idiosyncrasies in the activation of this knowledge modify accessibility, resulting in an incorrect annotation assignment. Persistent
slips
after a large number of examples may result from vigilance decrements in underlying attentional processes \citep{wiener_application_1987}. The classic serial position effect \citep{murdock1962serial} supports this distinction between knowledge-based mechanisms and attentional processes, in which early items are properly encoded and hence remembered while later items are only stored temporarily and subject to decay. Item order matters, particularly when the content to be acquired changes over time \citep{jacoby_proactive_2001}, as explained above under data challenges.
\subsection{Contributions}
This paper extends our prior conference publication \citep{pandey_modeling_2019}, with the following new contributions. 
\begin{itemize}
  \item[--] First, we present a generic human error framework
  of mistakes and slips, 
  adapted from the
  psychological theories that cover some common types of human errors and apply it to study human errors possible in an annotation task for streaming data, using the active learning system in a HITL-ML paradigm (Sections \ref{sec:3} and \ref{sec:4}). 
  \item[--] Second, we extend the validation of the proposed human error framework using a quantitative error model by presenting details of both lab-based and crowdsourcing-based testing experiments for the annotation task to filter relevant information from social media data streams collected during crises (Sections \ref{sec:5} and \ref{sec:6}). 
  \item[--] Third, we present a novel method for human error-mitigation in the active learning 
  paradigm for designing a stream processing system against several baselines (Section \ref{sec:7}). We also provide additional novel insights on different automated algorithmic approaches to prevent human error (Section \ref{sec:8}). 
\end{itemize}

The application of the proposed human error framework can be used to design Human-AI collaboration strategies and improve the performance of a human-in-the-loop approach for hybrid stream processing systems.

\section{Background}
\label{sec:2}
\subsection{Online active learning}
To the best of our knowledge, existing types of online active learning methods focus only on the possible machine/algorithmic errors. Prior literature (e.g., \cite{gama_survey_2014,almeida_adapting_2018}) provides extensive surveys of the different active learning paradigm-based methods. The primary categories include one group focused on a better sampling of the instance space for querying (e.g., addressing concept drift \citep{zliobaite_active_2014}), and another group focused on better learning of a discriminatory model. 

To improve sampling of the instance space, prior research has explored different mechanisms to drop the outdated/drifted class instances. The simplest way is to consider a fixed window over instance sequence and sample past instances from that window as they arrive. Windows can be specified by size and sampling on a first come first serve basis, or by time and sampling of instances from the last \textit{t} seconds/minutes/hours. These approaches do not represent well the characteristics of a data stream. Hence, alternative approaches were utilized in the past that uniformly sample and therefore, retain the characteristics of the underlying incoming stream of instances 
\citep{vitter_random_1985, ng_test_2008, yao_robust_2012, delany_case-based_2005, zliobaite_combining_2011, zhao_online_2011, salganicoff_density-adaptive_1993}. Other work does not completely drop all past instances but instead, reduce their weights for updating the classifier by an age-dependent factor 
\citep{koychev_gradual_2000, koychev_tracking_2002, helmbold_tracking_1994, klinkenberg_learning_2004, koren_collaborative_2010}.

To improve acquisition of the discriminatory model, prior research has mainly explored two strategies. The first is called the \emph{blind adaptation strategy}, which retrains the model without any detection of changes 
\citep{widmer_learning_1996, klinkenberg1998adaptive, klinkenberg2000detecting, lanquillon2001enhancing}. 
The other way of improving learning includes an informed strategy, which updates the model whenever a certain criterion is fulfilled like change detectors 
\citep{bifet_kalman_2006, hulten_mining_2001}. These criteria can also be aligned with the adaptation strategy \citep{gama_decision_2006, ikonomovska_learning_2011}, called model-integrated detectors. 

Our research premise is that the improvement of both types of the above active learning methods for stream processing systems require consideration of potential human annotation errors during the querying process as well, to be efficient and accurate in predictive model learning for the classifier. For simplicity, our method builds upon the blind adaptation strategy, which updates the model as we sample the instances in a sequence-based window. 

\subsection{Human annotation task and psychological processes}
Annotation quality can be affected by many factors. At the most basic level, a human annotation task can be conceptualized using signal detection theory (SDT) and its two fundamentally distinct parameters of discriminability (\textit{d'}) and decision criterion bias (\textit{beta}). Discriminability concerns the relationship between the mean signal strength of the distributions of positive and negative class instances. Nearly overlapping distributions pose difficult discrimination, such as using photographs to distinguish older from younger individuals that are close in age, whereas the overlap in signals is much smaller for gender discrimination from photographs \citep{nguyen_why_2014}. \textit{beta} in signal detection theory is an independent parameter, concerning the position of the decision criterion on these overlapping distributions, dropping it down to be more liberal to reduce the chance of misses (false negatives) or moving it up to be more conservative to reduce the chance of false alarms (false positives). The classic manipulation of \textit{beta} is achieved by the imbalanced distribution of positive and negative class instances or weighing the cost of misses and false alarms differently. 

Signal detection theory has been applied to the analysis of sequential industrial inspection tasks, resulting in the supposition of a vigilance decrements that affects judgment quality over time \citep{wiener_application_1987, mackworth_breakdown_1948}. This classic approach fails to recognize change over time in the relevant features in the data. Moreover, though influential in the perception literature, signal detection theory also fails to address several issues that arise in conceptual judgment tasks. Much later, \citet{kahneman_prospect_1979} elaborated a theory of bias to describe incoherence in decision making depending upon the influence of contexts such as prior belief or loss aversion. In this sense \emph{bias} is an umbrella term to characterize the systematic departure of decisions from rational analysis, which can account for a human annotator's errors given a drifting data stream. 

The annotation task is typically multi-class for a variety of applications that adds task complexity and hence, cognitive demand on the annotator. Following an initial training period, the failure to attain agreement between annotators on a multi-class coding scheme, known as inter-rater reliability in the social sciences \citep{creswell_qualitative_2016} has been generally attributed to a flawed coding scheme, rather than the cognitive challenge of learning the scheme and systematically applying it over time. 

Similarly, for information scientists developing machine learning models for data analytics, the appreciation of annotation as a psychological process emerges from the requirement for annotating large training datasets over an extended period of time, where each judgment matters. Although human annotation is often regarded as a gold standard, information scientists have noted that class imbalance leads to difficulties in appropriately representing the minority class to help human annotators learn the class concepts \citep{grant_evaluation_2017, broder_problematic_2017}. Information scientists have also observed that annotation styles affect human annotation quality to factors such as objectivity and descriptiveness \citep{cheng_how_2013}. Furthermore, annotation expertise affects quality, particularly in difficult tasks \citep{hansen_quality_2013}. Item position
with respect to its class concepts
(referred to as ``annotation schedule''), cognitive demand, and attentional processes may lead to annotation error \citep{burghardt_quantifying_2018}. Missing from both theory and method for human annotation tasks is a framework to organize and investigate specific human error types in the annotation tasks of hybrid stream processing systems.
Moreover,
unlike purely psychological research, the erroneous annotation of an individual item has consequences 
for
the machine learning model,
which learns
to automate the data annotation process. 
\begin{table}[]
\scriptsize
\begin{tabular}{p{0.22\textwidth}p{0.36\textwidth}p{0.31\textwidth}}
\Xhline{2.5\arrayrulewidth}
\multicolumn{1}{c}{\textbf{Type of Error}}                           & \multicolumn{1}{c}{\textbf{Potential Cause}}                                                                               & \multicolumn{1}{c}{\textbf{Mitigation Approach}}                 \rule{0pt}{3ex}                                 \\ \Xhline{2.5\arrayrulewidth}
\multicolumn{1}{|p{0.22\textwidth}|}{

\textit{*Mistakes induced by serial ordering}

}                   & \multicolumn{1}{p{0.36\textwidth}|}{
\begin{itemize}
  \item Concept not acquired yet 
\end{itemize}

}                                                                               & \multicolumn{1}{p{0.31\textwidth}|}{
\begin{itemize}
  \item Show frequent concept examples for learning, potentially informed by judicious selection such as near misses
\end{itemize}
}         \\ \hline
\multicolumn{1}{|p{0.22\textwidth}|}{\textit{*Slips induced by serial ordering}}                & \multicolumn{1}{p{0.36\textwidth}|}{
\begin{itemize}
  \item Imbalanced presence of a high-availability concept or a low-availability concept
\end{itemize}
}                                                   & \multicolumn{1}{p{0.31\textwidth}|}{
\begin{itemize}
  \item Limit extreme divergence from base rate for concept instances
\end{itemize}
}                                \\ \hline
\multicolumn{1}{|p{0.22\textwidth}|}{\textit{Mistakes and slips due to temporal and environmental constraints}} & \multicolumn{1}{p{0.36\textwidth}|}{
\begin{itemize}
  \item Concept memory decayed due to oversight in rapidly finishing the annotation task
  \item workload and stress of the external environment causing vigilance challenges in learning a concept
\end{itemize}
} & \multicolumn{1}{p{0.31\textwidth}|}{
\begin{itemize}
  \item Intervene reminders for concept examples
  \item Limit the number of concepts to annotate or the number of instances in a time unit
\end{itemize}

} \\ \hline
\end{tabular}
\caption{Framework of human annotation errors in hybrid stream processing applications. [*empirically studied in this article]}
\label{tab:framework}
\end{table}

\section{Human error framework}
\label{sec:3}
Our focus on a human error framework is intended to reveal the different human reasoning processes that result in erroneous annotations.
We assume a preliminary phase of the annotation task where instruction provides an initial understanding. However, this preliminary phase results in a mental representation of the concept (e.g., infrastructure damage during a disaster) at the beginning of an extended annotation task that is only partial, in the sense that the changing boundaries and nuances about a concept are learned while the annotations are performed. We also assume that the annotator can develop a mental representation of a concept by seeing a sufficient number of examples of this concept, even in the typical case where the examples are not annotated a priori. 

Following \citet{reason_human_2000}'s human error taxonomy built on  \citet{norman1981categorization}'s theory and broadly applied, including the analysis of medical domain errors \citep{zhang_cognitive_2004}, we distinguish two classes of errors for the human annotation task: mistakes and slips. Mistakes result from the absence of a correct cognitive representation of a concept. Slips are errors that happen despite acquiring the correct cognitive representation of a concept. 
Based on these broad classes, we present a 
framework
of human errors in the annotation task for stream processing in Table \ref{tab:framework} and explain the main error types below. We do not claim that all classes of human errors are equally prevalent or are equally consequential.

\subsection{Serial  ordering-induced  mistakes}  
The  annotation schedule  in which the tasks are presented to an annotator may prevent the annotator from adequately apprehending a concept, hence introducing mistakes. The main types of mistake include:
\begin{itemize}
  \item Concept not acquired yet: The annotator is asked to annotate an instance of a class for which s/he has not seen a sufficient number of examples to learn the concept overall. 
  \item Erroneous concept with missing or extraneous features: At best this blends categories and at worst creates uncertainty. 
\end{itemize}

\subsection{Serial ordering-induced slips} 
The annotation schedule in which judgments occur may cause slips, in which the annotator erroneously annotates an instance even if s/he has a correct representation of its concept. We identify two main cases for
the types of slips:
\begin{itemize}
  \item Slip favoring an available activated concept: In this case, metacognitive monitoring (vigilance) is suspended, resulting in an instance label that comes easily to his/her mind. Serial position, particularly the persisting activation of recent judgments, especially when reinforced with repetition has the potential to exacerbate slips that result in a false alarm (false positive). 
  \item Slip ignoring a minimally available concept: The complement of activation is effective inhibition. In this case, the correct category does not come to the annotator's mind, because its activation is too small compared to other categories. The annotator has not forgotten the concept, but it is inaccessible, resulting in the application of the available label instead of the correct one. This results in a miss (false negative). 
\end{itemize}

Because slips result from activation failures of fundamentally correct knowledge, concept training is unlikely to help. Both cases result from extreme local divergence from the base rate or the loss of metacognitive function, boredom, or fatigue. These can be addressed with proper annotation schedules.

Both types of errors described above, induced by primarily serial ordering constraints, are particularly vulnerable to a classification scheme that changes over time and underlying processes of proactive and retroactive inhibition on knowledge acquisition. As a whole, mistakes can be reduced by ordering instances to facilitate learning. This includes both a sufficient number of examples of each concept presented and reminders from old concepts, so that the annotator reinforces persistent and emergent critical distinctions between classes. Because the observable behavior (erroneous classification) is the same for both slips and mistakes, but the mitigation is different, the technical challenge is to identify the mechanism behind the observed error. 

\subsection{Other influences (temporal and environmental constraints) that induce mistakes and slips} 
As described in the background section, time and environmental constraints such as workload and its resulting stress during the annotation task can also cause human error. These constraints can cause vigilance and oversight challenges to the human annotators, causing slips and potentially, mistakes due to insufficient attention spent on the example instances to learn the concept. To limit the scope for the first study on such human annotation framework for stream processing applications, we do not consider such constraints in the experimentation and plan to explore these in future work. One of the future explorations to address such constraints include providing work specification, an amount of work, and a working environment that is appropriate, providing pauses to the worker, and so on.

In the following sections, we present three different experimental frameworks to reason about the existence of human errors and their mitigation by an algorithm: lab-based, crowdsourcing-based, and simulation-based. The lab-based error testing framework is similar to the conventional approach to experimentation in psychology, with greater control over the annotation task environment; however, a lab-based framework is difficult to scale to multiple annotators. The crowdsourcing-based approach can help to remedy the scalability challenge of the experimental setup. However, it provides less control on the setup to capture the annotators' behavior and their unacquired knowledge. Lastly, the simulation-based approach allows us to generate 
the streaming data samples, emulate human errors through an automated agent (referred `oracle'), and demonstrate error mitigation techniques.
However, it may oversimplify  observations of the real world and thus, could not capture the behaviors of all the different human annotators out there.

\section{Annotation task for hybrid stream processing systems}
\label{sec:4}
A hybrid stream processing application requires human annotation to adapt and improve the classification model continuously with new annotated instances. We define the specific annotation task for human error testing and mitigation to classify an instance from a given sequence/stream of Twitter instances (tweets) into \textit{k} classes.

We use labeled datasets from prior work in crisis informatics that contain labeled tweets related to natural disasters \citep{alam_crisismmd_2018}. We re-crawled the tweet instances from Twitter's API for acquiring metadata such as timestamp and discarded any tweets deleted since the data were originally collected. The three natural disasters include major natural hazards affecting Central and North America in 2017 – Hurricane Harvey, Hurricane Maria, and Hurricane Irma. The labels were created using a crowdsourcing platform, classifying instances into four major categories:
\begin{itemize}
  \item \textit{infrastructure and utility damage (c1):} information about any physical damage to infrastructure or utilities
  \item \textit{rescue, volunteering, and donation effort (c2):} information about offering help through volunteering efforts by a community of users
  \item \textit{affected individuals (c3):} information about the condition of the individuals during this disaster event
  \item \textit{not relevant or cannot judge (c4):} instance either does not contain any informative content or hard to decide.
\end{itemize}
We considered human labels with a confidence score (computed by the crowdsourcing platform for agreement between multiple annotators \citep{alam_crisismmd_2018}) greater than 65\% for ground truth 
labeled 
instances in our experimentation.

\section{Lab-scale annotation error testing}
\label{sec:5}
\begin{figure}[h!]
	\centering
	\begin{subfigure}[c]{0.70\linewidth}
		\includegraphics[width=\linewidth, trim={0cm 0 0cm 0}]{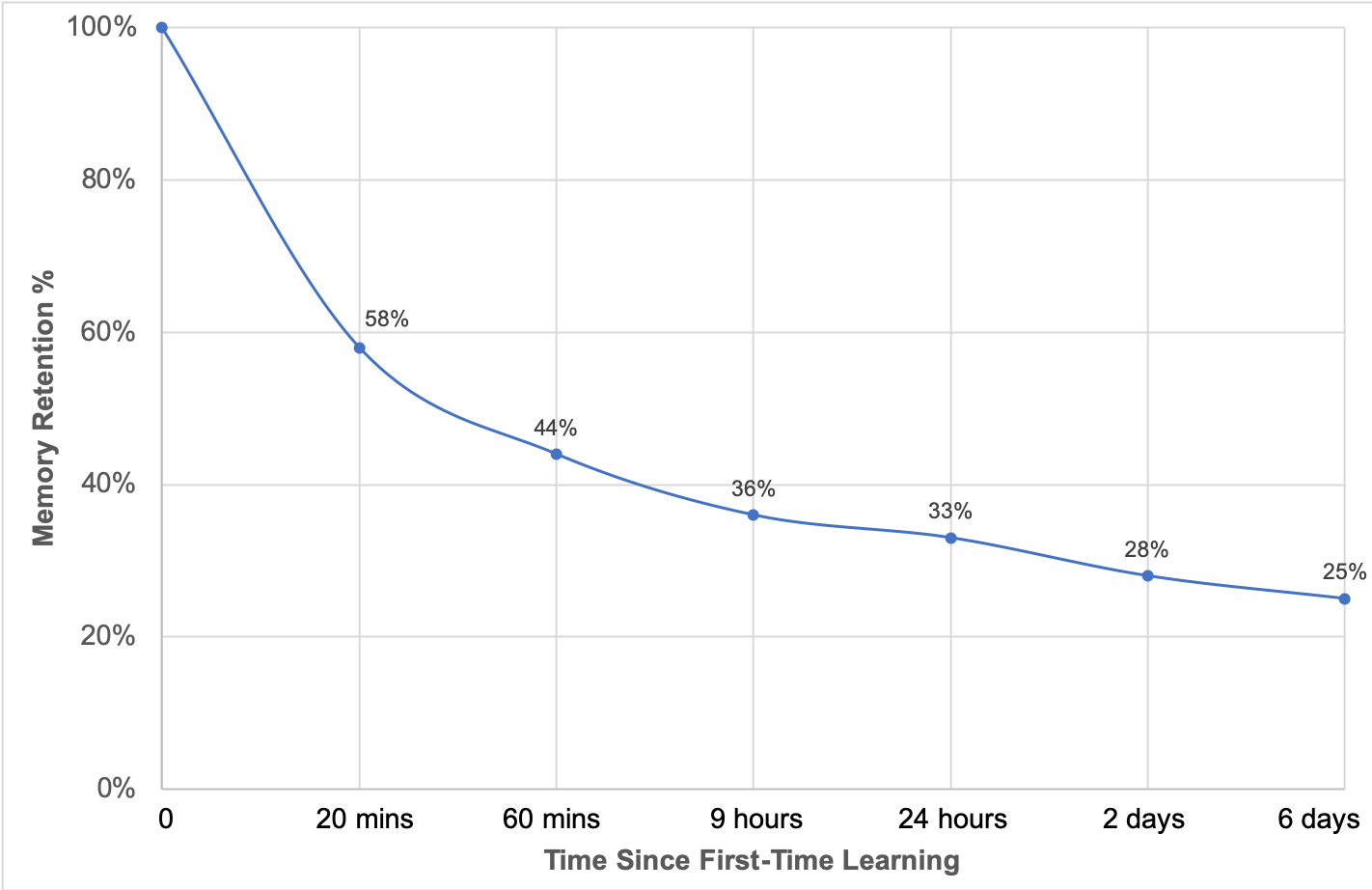}
	\end{subfigure}
  \caption{The effect of memory decay studied in Psychology \citep{ebbinghaus1913memory} 
  over time in learning or retaining conceptual knowledge. We investigate such effects of memory decay on the human annotation quality for hybrid stream processing systems and corresponding mitigation approaches.}
  \label{fig:memory}
\end{figure}
\subsection{Overview}
We focus on 
verifying the
effect of decayed memory behavior \citep{ebbinghaus1913memory}, 
which underlies the above-mentioned error types of serial ordering-induced mistakes \& slips and impacts the performance of both human annotation 
and
ML
models 
in the hybrid stream processing system. 

\subsubsection{Memory decay curve}
\label{subsubsec:memdecaycurve}
Psychologists have been studying memory-decay behavior in the context of learning and acquiring new knowledge for more than a century. The Ebbinghaus Curve – shown in Figure \ref{fig:memory} – is a fundamental and enduring contribution to the study of human memory. 
We observe from Figure \ref{fig:memory} an exponential decay of memory retention as the time since first learning passes. This exponential behavior has been widely observed in the psychology literature
\citep{brown_tests_1958, peterson_short-term_1959, melton_implications_1963}.
Moreover,
\citet{loftus1985evaluating} 
and
\citet{anderson1991reflections} 
have used an exponential function with respect to time to model memory decay or retention.
Hence, inspired
by the Ebbinghaus curve,
we model the $decaying\_score$ for the memory retention of an annotator for a particular class $(c)$. Specifically, we model the $decaying\_score$ for $c$ by observing how the annotator correctly annotates 
instances as an exponential function over time $t_c$ lapsed over its last seen annotated instance. We define the $decaying\_score(c)$ function in equation \ref{eqn:1} below:
\begin{equation}
\label{eqn:1}
  decaying\_score(c) \propto e^{-t_c}
\end{equation}

Moreover, to compute the probability of human annotators making an error, we use a function that is a vertical reflection of the aforementioned $decaying\_score(c)$ function along the \textit{x}-axis. For our experiments, we assume a parameterized \textit{sigmoid} function to compute the annotation error probability given the similar asymptotic nature of the vertical reflection of the exponential memory decay curves. We define the $error\_probability\_score$ function in 
equation \ref{eqn:2} below:
\begin{equation}
\label{eqn:2}
  error\_probability\_score(c) = \gamma \times \frac{1}{1 + e^{-\alpha t_c + \lambda}}
\end{equation}
Here the parameters $\alpha$, $\lambda$, and $\gamma$ represent different memory decaying intensities of human
annotators.
As each human 
annotator
has individual memory retention capability, the intensity of making errors in annotations 
varies
for
different human
annotators,
and hence these parameters help mimic different human memory decay behavior.
For 
verifying
the above function for human memory decay, we conducted a small-scale controlled lab study.
\subsection{Participants}
We selected three 
students working as Graduate Research 
Assistants
at an Information Technology research lab 
on social media research
to volunteer in this study. 
The participants included one female and two male students
(authors refrained from participation),
and all of them were in the age range of 25 -- 30. 
These students have been working with social media mining for more than six months. They have routinely participated in categorizing social media messages in the past. Hence, they
were well acquainted with social media messages during emergencies and were given a brief training session and clear instructions on annotating different class instances.
\subsection{Design}
The experiment used an annotation system for the annotation task defined in Section \ref{sec:4}. The input was a
sequence of tweet instances for the Hurricane Harvey disaster.
This synthetic input sequence contained instances with a random amount of irrelevant instances (noise) between ground truth annotated instances of any class to better observe human memory decay of the class and resulting errors. We added between 
one to four
irrelevant instances (randomly selected) between each of the ground truth annotated instances, and they were marked as ``not\_relevant\_or\_cant\_judge''. Our data stream contained 800 instances.
\subsection{Procedure}
All annotators were asked to annotate the instances into four class labels.
Three annotators labeled a given instance in the stream separately, with no ability to backtrack.
\subsection{Method}
Once we collected the three annotators' responses, 
we observed 
whether or not the annotators reveal memory decay effects for that class. 
Given the correct class concept for a given instance, we first store the time difference since we last observe any previous instance of that class concept. Moreover, we store the number of annotators who identified the correct class concept of that given instance. We plot the number of annotators who correctly identified the class concepts and the time difference for every instance.

\subsection{Results}
\begin{figure}[h!]
	\centering
	\begin{subfigure}[c]{0.68\linewidth}
		\includegraphics[width=\linewidth, trim={9cm 0cm 9cm 0cm}]{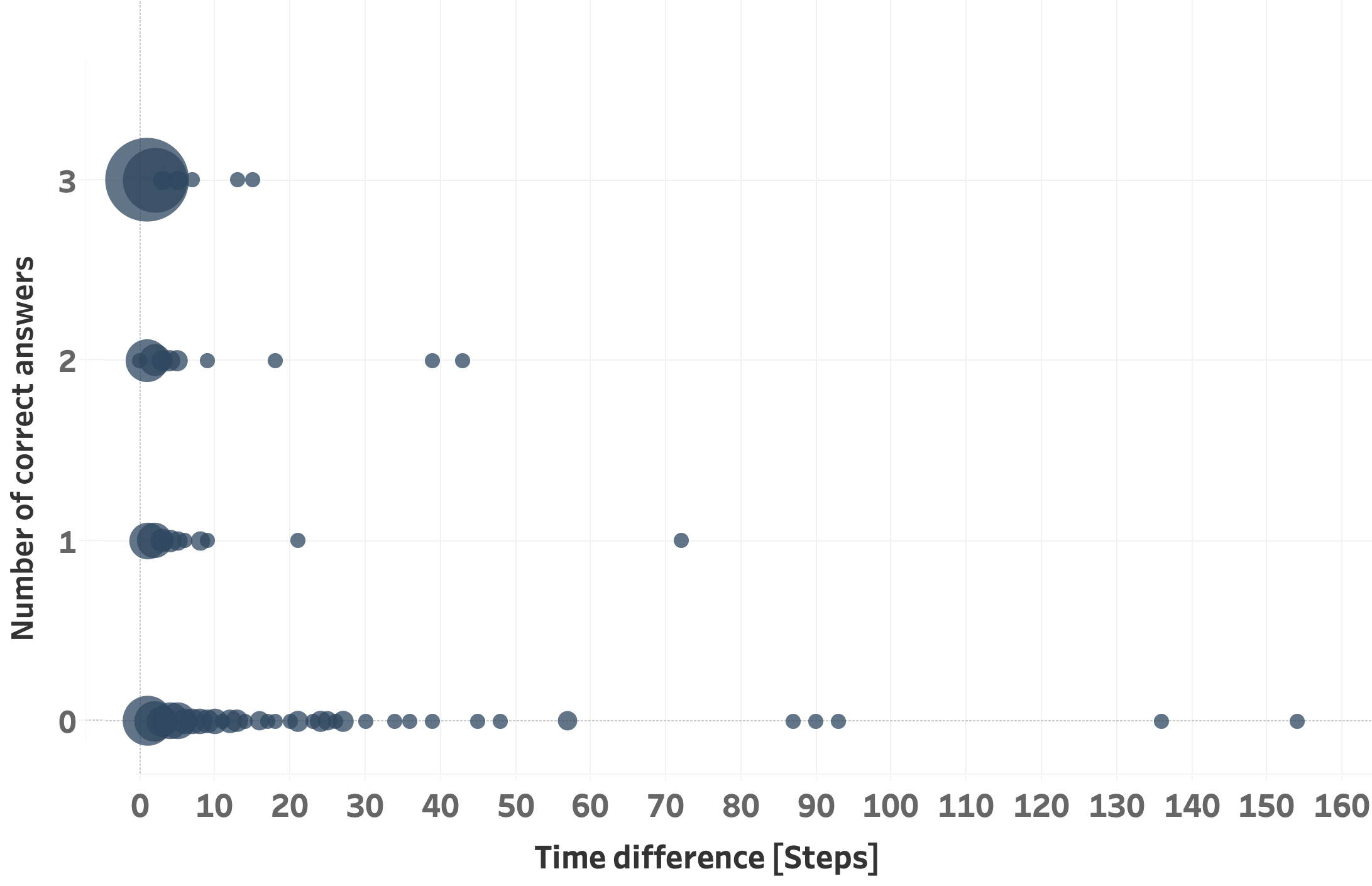}
	\end{subfigure}
  \caption{The temporal distribution of correct answers by human annotators shows a similar pattern as Figure \ref{fig:memory} for memory decay behavior. The probability of incorrect answers (error) increases for annotating a given instance in the sequence with increased steps between instances of the same concept.}
  \label{fig:exp_val}
\end{figure}
Figure \ref{fig:exp_val} shows the plot of how many instances of each class were correctly identified with respect to the time difference (in steps) between the appearance of consecutive instances of that class in the data stream. 
The size of the circle represents the number of instances that `y' annotators have correctly annotated with `x' time difference since they last observed that class instance in Figure \ref{fig:exp_val}. 
Due to the sequential nature of the experiment,
most of the class instances appear in very few ($<10$) steps,
and hence,
the figure is left-skewed. 
Moreover, we observe that many annotators incorrectly annotate the instances despite the class instances appearing 
frequently. This shows that multiple other influences can cause the annotators to make errors as described in Section \ref{sec:3}. However, we also
observe that when the time difference between the class concepts
increases, the chance of the annotators correctly identifying the class decreases exponentially and finally tends to zero correct annotation. In comparison, the highest chance of all the annotators picking the correct class in annotating an instance is when the time difference is close to zero.
\subsection{Discussion}
These results support the quantitative model of memory decay behavior as described in Section \ref{subsubsec:memdecaycurve},
verifying the exponential nature of memory decay 
(ref. equation \ref{eqn:1})
of annotators as they last see the instance of a particular class. 
Hence, we use an exponential function to compute the memory decay score for each class in our proposed \textit{Error-Avoidance Sampling} technique for error mitigation described in Section \ref{subsubsec:erroravoid}.
Moreover, as discussed in Section \ref{subsubsec:memdecaycurve}, a parameterized \textit{sigmoid}-based error probability function from equation \ref{eqn:2}
can be used
to induce an error in our algorithmic simulation experiments, later on, mimicking the real-world environment for a human annotation task in the stream processing systems.
We understand that the number of instances with large time steps was low. Hence, increasing the number of instances and the number of annotators would have shown more explicit exponential behavior of memory decay. Further, we observe a high inter-rater agreement ($0.82$ Cohen's Kappa score) between two of the three annotators, which is higher in comparison with the similar social media annotation task in the literature ($0.68$ Cohen's Kappa score from
\citet{ZHOU2021102554}),
but the third annotator had low inter-rater agreements with the other two ($0.48$ and $0.46$ Cohen's Kappa score respectively). We also observe that with the increase in time difference, none of the annotators could correctly identify a class. Within the limitation of the scalability of 
a
lab-scale study, we still achieve the same exponential decay behavior widely studied and suggested by the past psychology literature
\citep{loftus1985evaluating, anderson1991reflections}. 
Furthermore, our proposed error testing and mitigation approach can use any function, which closely resembles the vertical reflection 
properties of the exponential function and not just the \textit{sigmoid} function as an error-inducing function.

\section{Crowd-scale annotation error testing}
\label{sec:6}
\subsection{Overview}
Our crowdsourcing-based experiments seek to measure the prevalence of 
both mistakes and slips,
and the conditions under which these appear. The goal of these crowd-scale experiments is to motivate the design of algorithms seeking to
minimize these errors.
\subsection{Participants}
We asked
ten human judges  
to annotate six fixed sequences of instances per schedule for two schedules using the crowdsourcing platform.
 
\subsection{Design}
For the crowd-scale experiment,
we generated two types of annotation schedules for the task described in Section \ref{sec:4}, corresponding to mistakes and slips. For practical reasons of the cost and time of crowdsourcing, we limited the length of the schedules to 20 instances. For constructing the schedules, we used the labeled data as ground truth, and based on the labeled data distribution,
we chose the minority class c3 (instances about ``affected individuals'') as our target class for error analysis. 
The selection of c3 as the target class is used as an example to create a different annotation schedule 
because it was appearing the least in the data distributions, and hence, more prone to error.

For studying slips induced by serial ordering, we examine the case when instances of a target class (c3) are positioned with a mix of short and long gaps in the annotation schedule. 

We assume that non-uniform and infrequent occurrences cause the annotators to deactivate the knowledge of the target concept class, potentially leading to memory decay behavior. Thus, we 
hypothesize
that the annotation error per position of the target class instance should increase at the end of the annotation schedule
(H1).
Similarly, we study mistakes induced by serial ordering when instances of a target class (c3) are positioned with equal gaps in an annotation schedule. We observe the annotation error at each position in the schedule where an instance of the target class appears. We permute the instances of the target class on these positions. We hypothesize that uniform and frequent occurrences would allow the annotators to acquire the knowledge of the target class slowly. Thus, we hypothesize that the annotation error per position of the target class should reduce as we move toward the end of the annotation schedule
(H2).

The first annotation schedule corresponding to slip errors
(H1)
is \{c4, c1, c2, \textbf{\underline{c3}}, c1, \textbf{\underline{c3}}, c4, c1, c4, c1, c4, c2, c1, c4, c1, c2, c4, c2, c4, \textbf{\underline{c3}}\} and the second schedule corresponding to mistake errors 
(H2)
is \{c4, c1, c2, c1, c4, c2, c1, c4, \textbf{\underline{c3}}, c1, c2, c4, \textbf{\underline{c3}}, c1, c2, c1, \textbf{\underline{c3}}, c2, c4, c4\}. The underlined class label indicates the occurrence of target class instances and their position for analysis. For the three positions of the target class (c3) in an annotation schedule, we permuted the instances shown in those positions, leading to six experimental cases for each type of schedule.
\subsection{Procedure}
We used the Figure Eight platform (now called Appen) to obtain annotations for each type of experimental case.
Initially, each participant was given a set of example tweet messages for each class label to train them for the annotation. Next, they were asked to annotate 20 instances into the four labels described in Section \ref{sec:4}. 
For simplicity of the labeling process, we
separated the fourth label 
described in Section \ref{sec:4}
into ``not relevant'' and ``cannot judge'' respectively. We specified that the users should only look at the text while picking the label and not open any external link. The compensation amount for each task was \$0.15. Figure Eight platform uses several layers of protection to prevent inattentive workers. It encourages workers to maintain a high reputation within the system and removes anomalous workers, including automated responses (``bots''). They also indicated whether the response was tainted or not in any form, and we only included non-tainted responses.
After filtering all the tainted responses, we extracted
a total of 120 responses for the two schedules.
The reason for 
collecting many responses
from the crowdsourcing platform is to avoid the chances of other kinds of error influences such as inattentiveness, workload, and stress to overpower the cause of serial-ordering-based errors due to our proposed schedules.
\subsection{Method}
With the Figure Eight platform, we could not enforce participants to participate in only one experiment. Therefore, we
filtered out the responses from participants who have participated in multiple experiments. As a result, after filtering, we got 44 responses for the slips-based annotation schedule and 38 responses for the mistakes-based annotation schedule, respectively.
Once we filtered out the responses from 
duplicate
participants
(who participated in multiple experiments),
we 
analyzed the responses on both types of annotation schedules. 
We extracted annotations of the target positions for each response and checked if the annotators had incorrectly labeled to any class label other than \textbf{\underline{c3}}. If incorrectly labeled, we considered the error score as one, else zero. We accumulated error scores from all the responses made at first, second, and third positions, respectively. Next, we took the union of the errors made at the first and second positions and compared this with the third position's error 
using statistical significance testing.
\subsection{Results}


\begin{table}[]
\footnotesize
\begin{tabular}{p{0.07\textwidth}p{0.24\textwidth}p{0.13\textwidth}p{0.14\textwidth}p{0.13\textwidth}p{0.08\textwidth}}
\Xhline{2\arrayrulewidth}
Error Type & Potential Cause   & $1^{st}$ Position Error & $2^{nd}$ Position Error & $3^{rd}$ Position Error & \textit{p}-value \\
\Xhline{\arrayrulewidth}
Slip    & decayed knowledge  & 0.32        & 0.18        & 0.48        & 0.03  \\

Mistake  & unacquired knowledge & 0.13        & 0.25        & 0.18        & 0.07  \\
\Xhline{2\arrayrulewidth}
\end{tabular}
\caption{Annotation errors in the three positions of the target class instance (\textit{c.f.} description in Section~\ref{sec:6}), observed after 44 and 38 filtered crowdsourced responses on two types of annotation schedule respectively. The \textit{p}-value indicates the statistical significance for the difference between the error at the third studied position in comparison to the union of the errors at first and second positions.}
\label{tab:cs_annot}
\end{table}

Table \ref{tab:cs_annot} shows the results of crowdsourcing for the micro-average error rate (average error by an annotator for a given target instance) at the positions of the target class in the schedule. 
We note inverse functions for the position effect on the potential knowledge acquisition and human error, depending upon the manipulation. 
The high micro-average error rate at the third position for slip error type supports the distinction with respect to the placement of the target instance.
We also ran a 
paired
two-tailed 
t-test
to observe any statistically significant difference between the errors of the specific positions. 
For the slip errors-based annotation schedule (H1), we observe $t(43) = - 2.20$ with \textit{p}-value of $0.03 < 0.05$, whereas we observe $t(37) = 1.87$ with \textit{p}-value of $0.07 > 0.05$ 
for the mistake errors-based annotation schedule (H2). 

\subsection{Discussion}
From the paired two-tailed
\textit{t}-test
results mentioned above,
we found a significant difference (\textit{p}-value =
0.03)
for the error between the last position and the average of the earlier two positions in a sequence that depicts slips due to potentially memory decay of the knowledge of the target class. This shows that a significant gap with no occurrences of a class indeed increases annotation errors for that class and suggests frequent reminders of the concept/class are needed in a sequence for annotation tasks,
and hence, the hypothesis
H1
made for slips error was accepted. Whereas, 
with
due to the low degree of freedom ($37$) and \textit{p}-value of $0.07 > 0.05$, we cannot accept or reject the hypothesis H2 as
we cannot rule out a difference between the error at the last position and the average of the earlier two positions in the case of some annotations (\textit{p}-value of
0.07). 
However, it is a major challenge to model the acquisition of knowledge for a concept due to a lack of information on the prior knowledge or experience level of the annotators,
and hence, we could not validate the hypothesis made for the error type  \emph{mistakes}. We discuss these limitations in Section \ref{sec:8}.
Motivated by these promising results, we next describe large-scale simulations and algorithmic solutions to mitigate such human errors in the HITL-ML paradigm-based stream processing applications focusing on slips due to potential memory decay.

\section{Simulation-based error testing and mitigation}
\label{sec:7}
\begin{figure}[h!]
  \centering
  \begin{subfigure}[c]{\linewidth}
    \includegraphics[width=\linewidth, trim={0.7cm 0 2.7cm 0}]{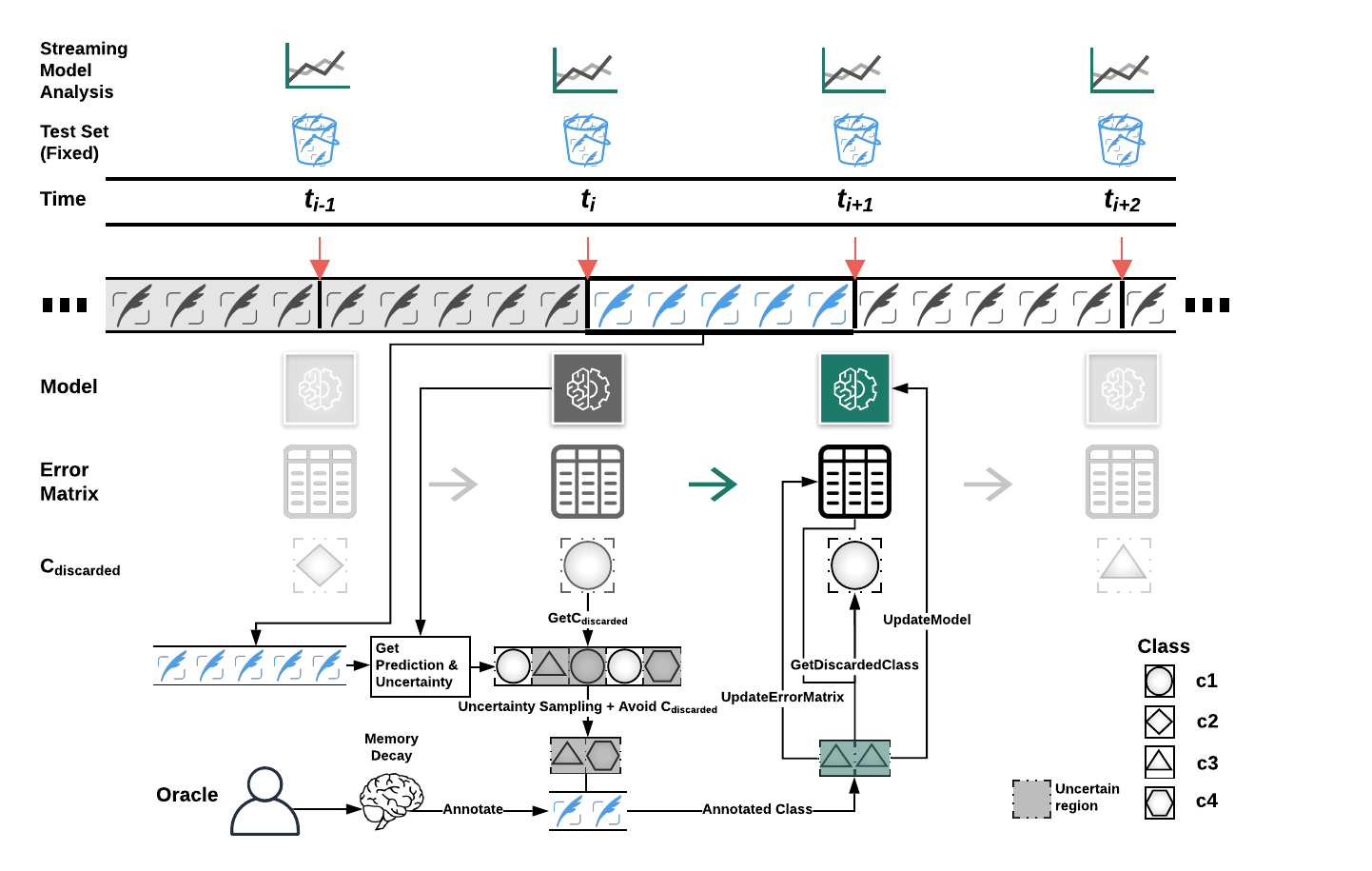}
  \end{subfigure}
  \caption{Summary of the proposed Error-Avoidance Sampling-based human error mitigation algorithm. At every interval, the streaming model predicts the potential class labels with uncertainty. The uncertain instances and instances of the discarded class label are filtered out, and the remaining are annotated by the oracle, which mimics the memory decay of human annotators. Based on the new annotations, the streaming model is updated, and the next class concept to discard is computed to avoid human errors in the annotation.}
  \label{fig:algo}
\end{figure}
We simulate the annotation task in an active learning paradigm for online stream processing \citep{zliobaite_active_2014}. We design a novel method for generating a dynamic annotation schedule (instance sampling and ordering) for an annotator (simulated ``oracle'') such that the schedule attempts to minimize human errors
(Serial Ordering-induced Slips)
and maximize the overall performance of the active learning paradigm. 

\subsection{Mitigation algorithms}
Our method first samples a batch of $m$ instances from a time interval $[t_i,t_{i+1})$ by using a conventional uncertainty sampling algorithm for an active learning paradigm, followed by applying constraints to select only $n$ $(n < m)$ instances for annotation that minimize the potential human memory decaying error, and then, update the machine learning model for predictions in the next time interval $[t_{i+1},t_{i+2})$. For annotations by ``oracle'' in the simulation, we use ground truth labels (c.f. Section \ref{sec:4}) 
along with the memory decay to simulate human errors (explained later in Section \ref{subsec:al}). 
We propose three types of algorithms (the first two being the baselines) based on diverse sampling strategies for selecting instances to annotate at the end of time interval $[t_i,t_{i+1})$:
\subsubsection{(Baseline) Algorithm 1: random sampling}
We randomly sample $n$ instances from the batch of $m$ streamed instances in the recent interval of $[t_i, t_{i+1})$. We hypothesize that random sampling can address the issue of data distribution changes for concept drift by selecting an instance from any region in the concept space, although it may be inefficient to improve learning performance over time. For consistency, we use an equal number of samples for this algorithm to the number of instances sampled by the popular active learning paradigm of uncertainty sampling, as described next.
\subsubsection{(Baseline) Algorithm 2: uncertainty sampling}
We predict the classes of the incoming batch of instances with the current active learning algorithmic model. At the start of the time interval of $[t_i,t_{i+1})$, along with the new incoming instances, we also receive a model, which was trained with all the annotated instances before $t_i$. We use this model for prediction. After prediction, we select the classified instances with uncertainty in the prediction confidence -- probability in the range of $[30\%, 70\%]$. We provide the uncertainty region instances to the oracle and obtain its annotations. We hypothesize that the model will become more robust if it starts learning from the cases on the decision boundary region \citep{winston1984artificial}. 
\subsubsection{(Proposed) Algorithm 3: error-avoidance sampling}
\label{subsubsec:erroravoid}
This algorithm relies on uncertainty sampling to first select candidate instances from uncertain regions. It then discards the instances whose predicted class (from the model received at time $t_{i}$) could either add noise to the new model or tend to be forgotten by the oracle (i.e., memory decay in human learning behavior toward that class). The algorithmic flow is formally described in Figure \ref{fig:algo}. 

Specifically, at the beginning of each interval $[t_i,t_{i+1})$, we receive four information components described below: 
\begin{itemize}
  \item[--] \textbf{Incoming Instances.} All incoming streams of instances at time interval $t_i$ shown by a string of tweet symbols in Figure \ref{fig:algo}.
  \item[--] \textbf{Model.} Streaming active learning model, which is trained with all the instances that are annotated by the oracle before $t_i$.
  \item[--] \textbf{Error Matrix.} This is a matrix that contains information about each instance annotated by the oracle from the previous two intervals (i.e., $[t_{i-2},t_{i-1})$ and $[t_{i-1},t_i)$) and the current interval (i.e., $[t_i,t_{i+1})$). Each row in the matrix represents an annotated instance and contains information about its arrival time, annotated class by the oracle, and the set of per-class prediction error. 
  The per-class prediction error for a class is the average error of predictions for the annotated instances of the class present in the current error matrix, and it is computed using the active learning model updated with the current instance. 
  For example, if an instance $X$ has been annotated with class $c3$ by the oracle at time $t_X$, we store the values $X$, $t_X$, and $c3$ to the error matrix. In addition, we also have column information for per-class prediction error as 
  $E(c_i | c_j)$ where 
  $i \epsilon [1,num\_class]$, $j \epsilon [1,num\_class]$, and $i \neq j$. 
  In our case $num\_class = 4$. 
  %
  In this example, all values of $E(c_i| c_j)$ for $i \epsilon [1,4]$ and $j \epsilon \{1,2,4\}$ are copied from the previous row to the current row for time $t_X$, as the current instance that has arrived is annotated with class $c3$ (i.e., $j=3$). Next we calculate $E(c_i|c3)$ for $i \epsilon [1,4]$. 
  
  For computing $E(c_i|c3)$ at time $t_X$, we predict all previous instances in the error matrix with the updated active learning model (re-trained with the annotated instance $X$). Next, we calculate the per-class F-measure ($F\_{measure}(c_i)$) where the annotated class is considered as true class to compare with the predicted class, and thus, we compute per-class prediction error ($1 - F\_{measure}(c_i)$). Finally, we get the per-class prediction error $E(c_i|c3)$ for each class $c_i$ 
  in the final row of the error matrix for the instance $X$. This error matrix helps in determining a class to discard 
  as explained next.
  
\item[--] \textbf{Discarded Class ($C_{discarded}$).} This represents the class that has induced the most errors to the other classes or whose instances appear very frequently, causing memory decay for instances of other classes. To identify the discarded class $C_{discarded}$, we first compute the \textit{error avoidance score} and \textit{decay score} for each class to get its final score (explained below) and then, choose the class which has the highest final score. 
\end{itemize}

\noindent \textit{Algorithm steps.} We now go through the flow of our proposed algorithm summarized in Figure \ref{fig:algo} and refer to corresponding functions and variables in parenthesis. First, for each instance $X$ (represented by a tweet symbol in Figure \ref{fig:algo}), we predict its class based on our current model received at $t_i$. Second, we select the instances that are in the uncertain region (dark-colored tweet symbols) and are not predicted with the class that is $C_{discarded}$. We believe that at each interval, $C_{discarded}$ represents a class whose instances cause error in the active learning model. Third, we schedule the selected instances for annotation by the oracle and update our model ($UpdateModel$ function). Finally, we update the error matrix by adding a row for instance $X$, and storing values as per the error matrix definition. 

To decide which class to discard, we compute two scores: \textit{error avoidance score} and \textit{decay score}. 
The error avoidance score determines the total error induced in the model for other classes due to the addition of the current instance into its training set. While the decay score determines the class that appears with excessive frequency in the stream, causing memory decay for other classes and thus, leads to annotation error. Note that we use the error matrix to decide the classes to discard only after the first three intervals. 

We calculate the error avoidance score for each class $c_j$ ($j \epsilon [1,4]$) as:
\begin{equation}
  GetErrorAvoidanceScore(c_j) = \sum_{k = 0}^{m} \sum_{i = 0}^{n} E_{k, (c_i, c_j)}
\end{equation}
where \textit{k} is the total number of instances in the error matrix. 

Next, we calculate the memory decay score to determine which class appears too frequently in the stream. For each class $c_j$, we calculate the score as: 
\begin{equation}
\label{eqn:4}
  GetDecayScore(c_j) = e^{-\Delta T_j}
\end{equation}
where $\Delta T_j$ is the time difference from the recent two occurrences of the instances of class $c_j$ in the error matrix. 
Note that the $GetDecayScore$ function is a simplified version of the decaying score function defined in equation \ref{eqn:1}. We have not parametrized this function as we generalize this decaying factor the same for all humans. Hence the memory decay score only depends on the time since the last viewed class instance irrespective of the different decaying intensities of different humans. Additionally, we observe that the memory decay score for any class is reset every time that class instance gets picked for annotation. Moreover, as the time difference increases, the decay score decreases exponentially similar to the behavior of memory decay in psychology, which has been 
discussed in
the lab-scale 
experiment
in Section \ref{sec:5}.

Lastly, the final score for each class $c_j$ is defined as:
\begin{equation}
  Score_{c_j} = GetErrorAvoidanceScore(c_j) \times GetDecayScore(c_j)
\end{equation}

Once we calculate the final score for each class, we determine the class $c_j$ with the highest score as the error-inducing class to discard ($GetDiscardedClass$ function). 
We observe that while choosing the discarded class at each interval, we are not only looking for the memory decay factor but also the input text of the instance and the effect it has on the model performance for other class labels.

\subsection{Simulation experiments}
We describe the data preparation for the simulated stream processing task and the active learning paradigm.

\paragraph{Data preparation} We use labeled datasets from three major hurricanes in Central and North America as described in Section \ref{sec:4}. We split the data into training, test, and warm-up sets. Twenty percent of the whole dataset is used as a test set. From the remaining 80\% of the data, we randomly picked $n$ instances ($n = 20$) of each class to create a warm-up set; the rest constitutes the training set. As we have a class imbalance in our data, we use an equal number of instances across classes for creating our warm-up phase model for robustness. The training data are sorted based on the arrival time of an instance (tweet) in the stream. After sorting, we divided the data into equal bins of size $N$. At each interval, $N$ instances would arrive for annotation and get filtered for inclusion in the training set based on our mitigation algorithms. 

We fix $N$ based on volume as our labeled dataset is not continuous in a real-time setting but is distributed over an extended period, given it was annotated through a crowdsourcing method in prior work. Hence, we cannot fix $N$ based on time units (seconds, minutes, etc.), but our approach is generic and applicable for other scenarios. 

\subsection{Active learning environment}
\label{subsec:al}
We implement the active learning paradigm following previous work \citep{zliobaite_active_2014}. First, we train the base model with the warm-up set and then, keep updating with the new incoming instances sampled by our baseline or proposed algorithms. 

We used a fairly standard set-up for text classification, using pretrained GloVe-Twitter embeddings \citep{pennington_glove_2014} with 200 dimensions for generating word-level features and then averaging the word-level embeddings to represent tweet-level features. We train a linear SVM model and measure the performance on the fixed test set.

For every interval $t_i$, we receive $N$ instances for seeking the annotator feedback to acquire more labeled data for retraining the current model. Depending upon the mitigation algorithms, i.e. Random, Uncertainty, or Error-Avoidance Sampling; we sample the instances to obtain annotations from the oracle. To mimic human behavior, the label for the instance given by the oracle annotator is not always correct. Based on the lab-scale experiment's
discussion
in Section \ref{sec:5}, we note that the
error probability score for annotators making errors in annotation can be depicted using a parameterized sigmoid function
in our annotation task. Thus, we utilize the value of a sigmoid function with different parameters to find the probability that the oracle generates a correct or erroneous label due to memory decay of the class as given in the formulation of equation
\ref{eqn:2}.
We define the ``Memory Decay'' component in Figure \ref{fig:algo} 
precisely to the oracle to highlight this memory decay behavior of the oracle. We use 3 different parameter settings to add errors through the memory decay behavior of the oracle (annotator): 

\begin{enumerate}[label=\arabic*)]
  \item \textbf{Slow decaying:} computes a sigmoid function with parameters estimated from errors observed in the crowd experiment: $\alpha = 0.0434$, $\lambda = 0.9025$, and $\gamma = 0.75$.
  \item \textbf{Fast decaying:} uses a sigmoid function that converges to 1 faster than the slow decaying and induces errors more frequently: $\alpha = 0.03$, $\lambda = 1.00$, and $\gamma = 1.00$.
  \item \textbf{No decaying:} assumes that our oracle always gives the correct labels and does not have any memory decay of the knowledge of any class. Hence, we use the true ground truth labels for each annotation. 
\end{enumerate}

\subsection{Results}
\begin{figure*}[h]
 \centering
  \subfloat[]{\includegraphics[width = 1.5in,trim={0.7cm 0cm 1.6cm 0cm},clip]{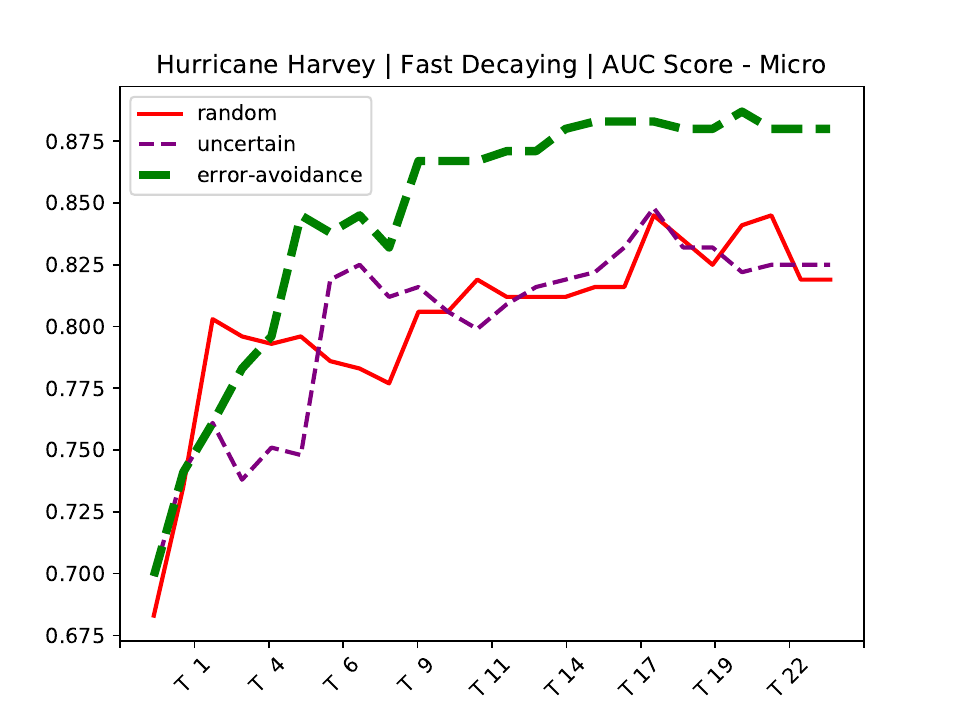}}\hfill 
  \subfloat[]{\includegraphics[width = 1.5in,trim={0.7cm 0cm 1.6cm 0cm},clip]{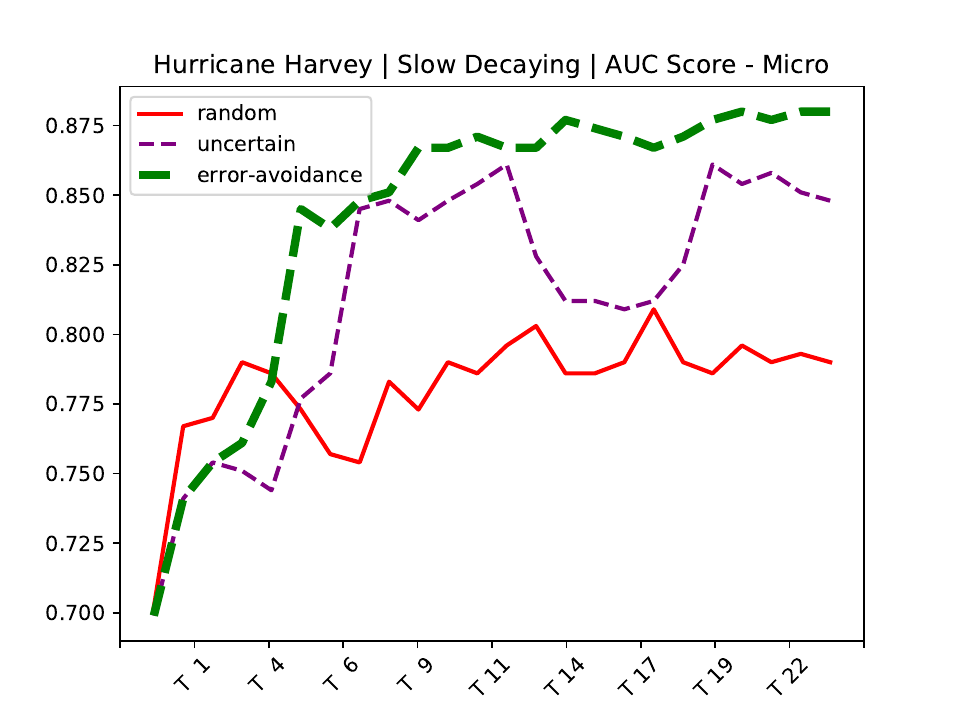}}\hfill 
  \subfloat[]{\includegraphics[width = 1.5in,trim={0.7cm 0cm 1.6cm 0cm},clip]{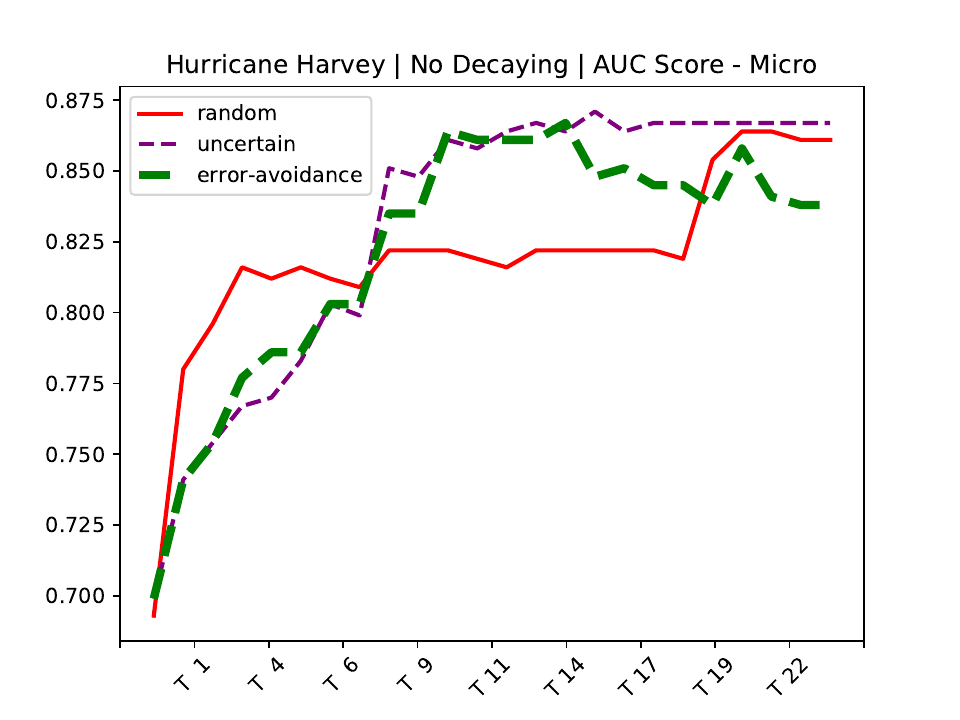}}
 \\
  \subfloat[]{\includegraphics[width = 1.5in,trim={0.7cm 0cm 1.6cm 0cm},clip]{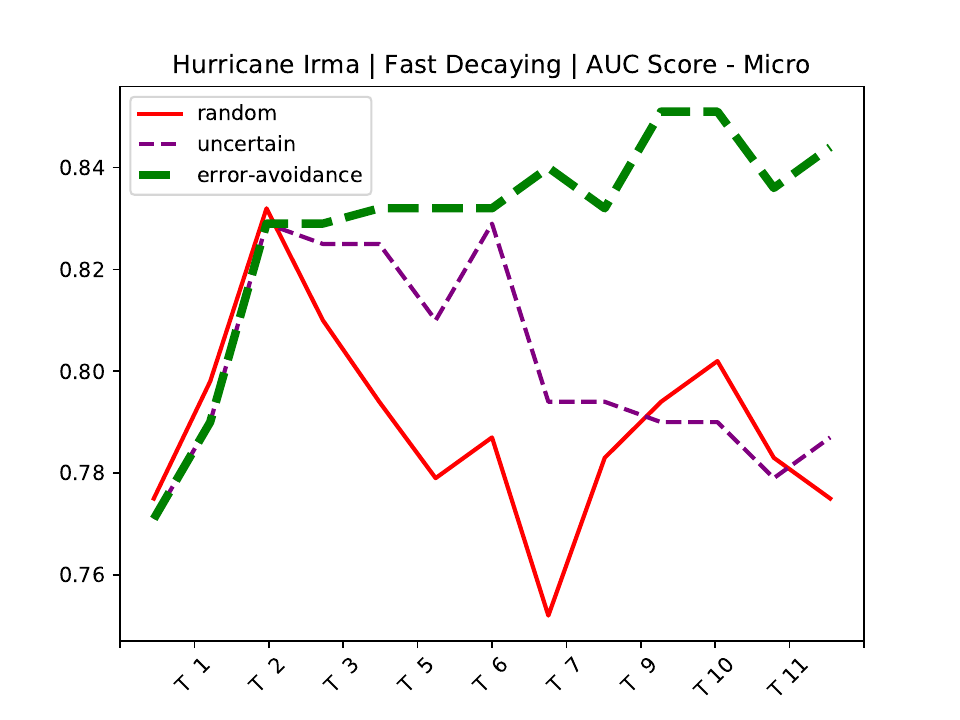}}\hfill 
  \subfloat[]{\includegraphics[width = 1.5in,trim={0.7cm 0cm 1.6cm 0cm},clip]{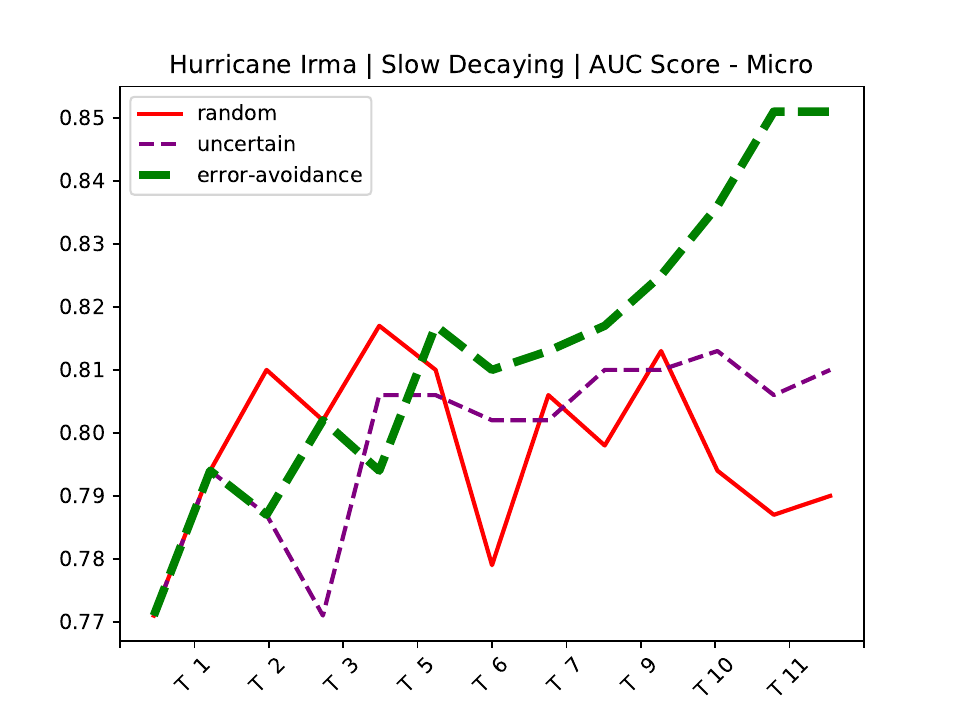}}\hfill 
  \subfloat[]{\includegraphics[width = 1.5in,trim={0.7cm 0cm 1.6cm 0cm},clip]{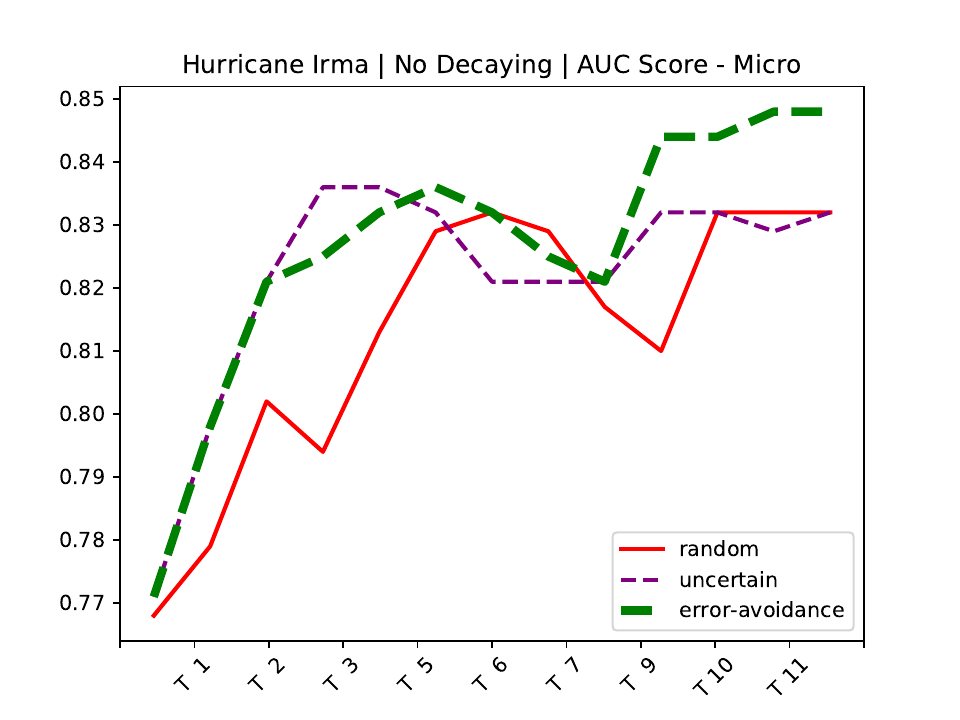}}
 \\
  \subfloat[]{\includegraphics[width = 1.5in,trim={0.7cm 0cm 1.6cm 0cm},clip]{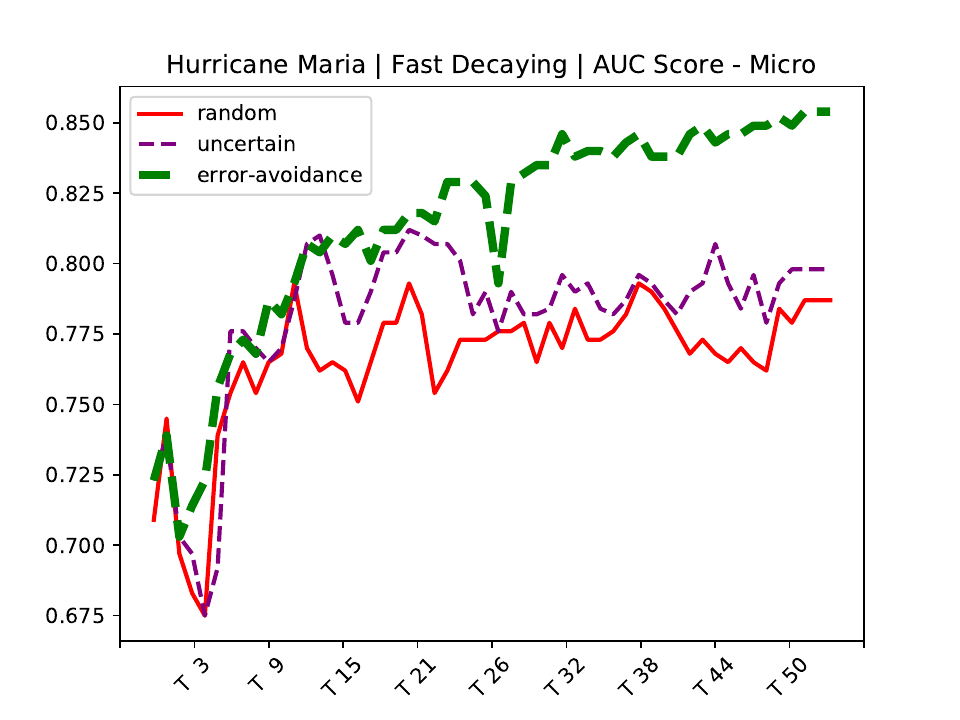}}\hfill 
  \subfloat[]{\includegraphics[width = 1.5in,trim={0.7cm 0cm 1.6cm 0cm},clip]{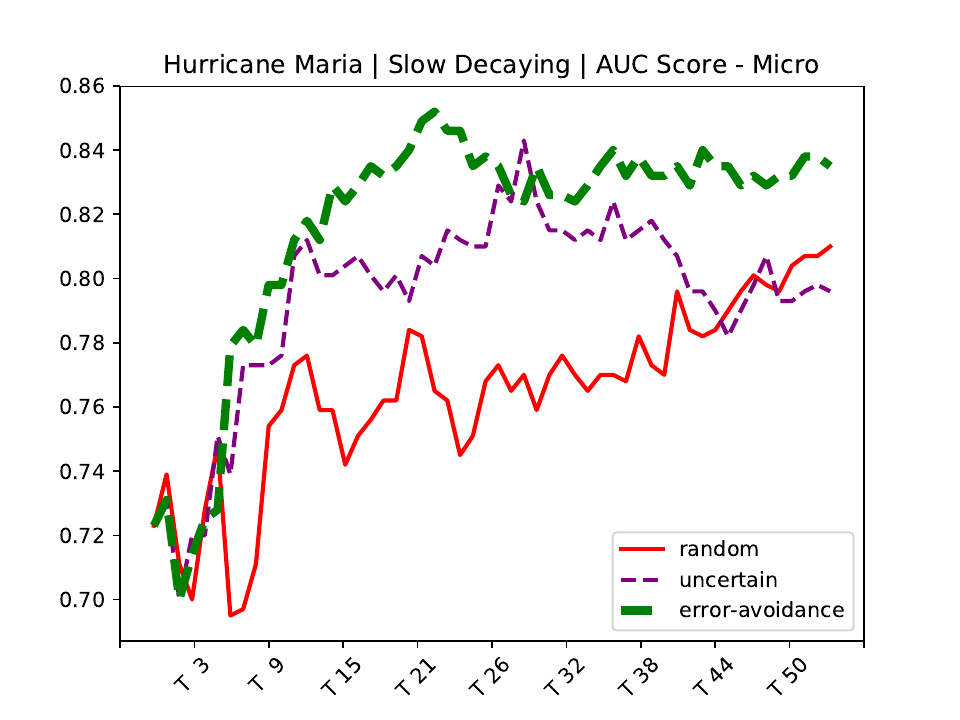}}\hfill 
  \subfloat[]{\includegraphics[width = 1.5in,trim={0.7cm 0cm 1.6cm 0cm},clip]{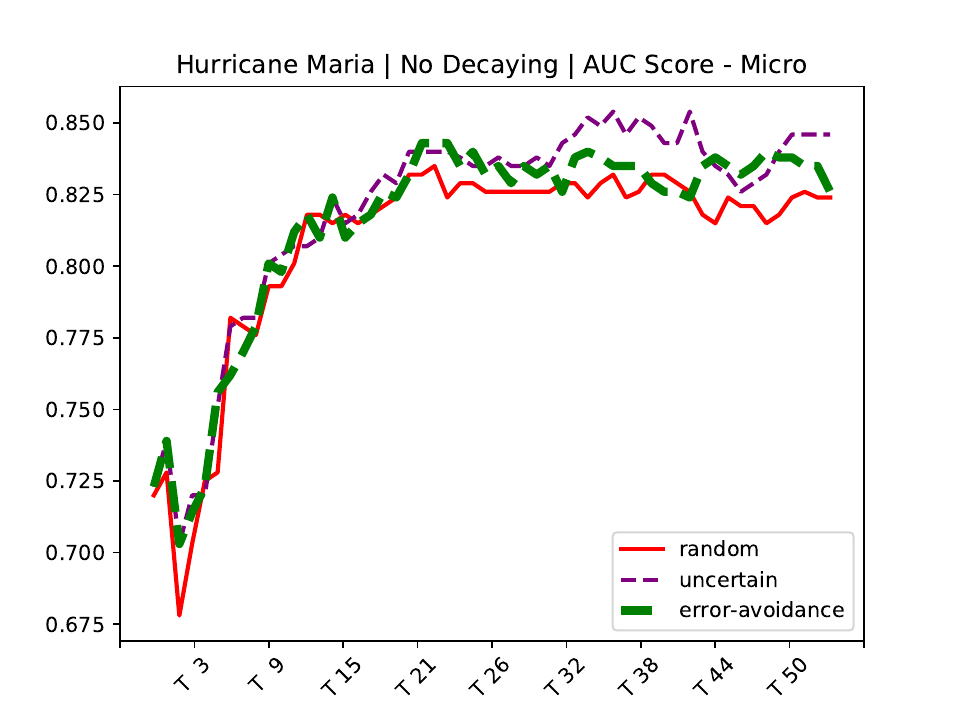}}
\caption{
AUC score 
of mitigation algorithms 
for three hurricane datasets by 
various decay settings, showing superior performance of the proposed error-avoidance sampling in the case of memory decay errors.
}
\label{fig:results}
\vskip -0.1in
\end{figure*}

We experimented across three event datasets for a robust evaluation of our simulation algorithms. These event datasets have a varying number of instances per interval: Hurricane Harvey has $N = 36$, Irma has $N = 59$, and Maria has $N = 18$. Therefore, our results have taken into account the different burstiness of the streaming data instances during the real disaster event. We report the AUC scores for every experiment on the fixed test set per event. Figure \ref{fig:results} shows the AUC scores of our three mitigation algorithms using different decay behavior settings of the oracle annotator, across all three datasets. We computed the micro average of the AUC scores at each time interval for different mitigation settings as the model is trained differently on each of them. The behaviors of accuracy and F-measure follow a similar pattern to those of AUC, thus, we omit those figures for brevity;
however, they are included in the supplementary materials for transparency.
The results demonstrate the effectiveness of our proposed annotation scheduling approach in contrast to the two baselines for mitigating annotation errors, and thus, improve the automatic classification performance.

\subsection{Discussion}
We mimicked a real-world annotation scenario by inducing different types of memory decay-based human errors (slow vs. fast decaying) in a simulated annotation schedule. The error mitigation algorithm based on our error-avoidance sampling technique can select instances for a human to annotate, which mitigates the effect of human memory decay and improves AUC scores over time across all the event datasets, despite varying numbers of instances per interval. Also, for the first three intervals, both the simple uncertainty-based and error-avoidance sampling-based algorithms are equivalent in performance during the initial time. This is consistent with an interpretation in which our algorithm may still be learning about the class that induces errors to other classes or make other classes forgotten by the annotator. Both of these algorithms show a gradual increment of performance as the new instances arrive, compared to the random sampling algorithm with highly variant behavior of the learning model. These observations support the claim that our proposed algorithm could help improve active learning paradigm-based real-time systems. 

In the case of a no-decaying simulation setting, where the oracle always (but unrealistically) provides the correct label, all mitigation algorithms perform similarly to each other. This is possible due to similarity in frequency and the amount of correct oracle feedback, which constantly updates the model with new training data that gradually improves on the test set.

In the case of our error-avoidance sampling-based algorithm, the chances of inducing human error decrease due to accounting for the likelihood of memory decay of the knowledge about a class, which attempts to reduce the expected errors in annotating instances of all classes. 
Through the $GetDecayScore$ function in Equation \ref{eqn:4}, we give more weight to the class to discard whose instances are appearing more often in the streaming data. Hence, those class instances will be discarded in the next period so that the annotators do not forget the other classes that are appearing less frequently. Moreover, as the $GetDecayScore$ function is independent in our proposed algorithm, it can take any memory decay function making our error-avoidance sampling-based algorithm generic.

In summary, our study of human error types in the annotation of streaming data presents novel insights on their effect on the performance of active learning systems in the (HITL-ML) paradigm-based stream processing systems. 
This study raises awareness of 
instance ordering when designing crowdsourcing-based tasks. In particular, we recommend using an annotation schedule in an active learning paradigm-based system that can help reduce human errors.
To the best of our knowledge, this is the first study to investigate a principled framework for quantifying human annotation errors for social stream processing and develop mitigation methods by better understanding the human annotation task as a psychological process.

\section{Conclusions}
\label{sec:8}

We defined a framework of human errors, including mistakes and slips in the context of stream processing, based on psychological theories of human errors. We specifically focused on a quantitative model of memory decay behavior in the context of annotation tasks of humans, given that it is a common cause for both mistakes and slips. We validated the existence of memory decay-based annotation errors in a variety of experimental setups from lab-scale to crowdsourcing and provided evidence for the conceptual distinction between slips and mistakes for stream processing applications. We performed simulation-based studies to test a novel error mitigation algorithm targeted to slips that minimizes the likelihood of memory decay
in an active learning paradigm-based human annotation task. 
The proposed method for human error mitigation can help design Human-AI collaboration systems for efficient stream processing for social media and web data in general. Such systems would require not only fewer human annotations but also reduce errors and decrease annotator memory decay. 

\paragraph{Limitations and future work.} We have provided a proof-of-concept using an over-simplified model of human memory 
\citep{anderson2000learning}. In particular, we have simplified the activation and decay functions and the self-reinforcing effect of classification on persisting knowledge of class concepts in an annotation task. 
Different error probability score functions other than sigmoid may better model
the memory decay behavior of humans.
Our approach to the characterization of human annotation error is also focused on cognition and ignorant of exogenous influences on cognition, including the physical and social environment \citep{hollnagel_cognitive_1998}. We do not claim that this study covers all types of human annotation errors in stream processing, in particular, the knowledge modification problem posed by changes in streaming content. In focusing on serial effects, we have ignored the effect of absolute time. 

Our small crowd-scale study provided insufficient power to definitively distinguish the mistake-based error.
A future large-scale crowdsourced study could provide more definitive support for the effectiveness of our proposed error mitigation algorithm. 
Nevertheless, we have documented that blind confidence in human annotation as a gold standard is gravely erroneous. 
Dramatic performance improvement results when the annotation is utilized with an appreciation for the human processes that generated it and might lead to errors. We hope that our framework provides a foundation for studying diverse types of annotation errors and causes, beyond text 
to image object recognition 
for a variety of stream processing applications, such as addressing burnout or inattentive worker errors in the future for human-AI teaming. 

\paragraph{Reproducibility} Human annotations and code implementations are available upon request for research purposes.
\paragraph{Supplementary material} Supplementary material associated with this article can be found, in the online version, at \href{https://doi.org/10.1016/j.ijhcs.2022.102772}{10.1016/j.ijhcs.2022.102772}.

\section*{Acknowledgment}
Purohit thanks U.S. National Science Foundation grant awards IIS-1657379 and 1815459, and Castillo thanks the funding received from the ``la Caixa'' Foundation (ID 100010434), under the agreement LCF/PR/PR16/51110009 by the HUMAINT program (Human Behaviour and Machine Intelligence), Joint Research Centre, European Commission for partial support to this research. The opinions expressed are those of the authors and do not reflect those of the sponsors.


\bibliography{references_ijhcs}

\end{document}